\documentclass[10pt,leqno]{amsart}
\usepackage{graphicx}
\usepackage{indentfirst,csquotes}
\usepackage{booktabs,tabularx,array}

\usepackage{amssymb,amsthm,amsmath}
\usepackage{xcolor,paralist,fancyhdr,etoolbox}

\usepackage{lipsum}

\makeatletter
\newcommand{\affiliations}[1]{\gdef\@affiliations{#1}}
\let\@affiliations\@empty
\apptocmd{\@setauthors}{%
  \ifx\@affiliations\@empty\else
    \par\vspace{0.5\baselineskip}%
    {\centering\footnotesize\itshape \@affiliations\par}%
  \fi
}{}{\PackageWarning{affil}{Failed to patch \string\@setauthors}}
\makeatother

\usepackage{natbib}
\usepackage{hyperref} 
\usepackage{caption}

\hypersetup{ colorlinks=true, linkcolor=black, filecolor=black, urlcolor=black }

\usepackage{lipsum}

\begin{document}
\title[]{Uncovering Non-Normality in Information Flow: Network Structure and Dynamics of Social Media Cascades}

% Qianyun Wu, Bruno T. Sugano, Genta Toya, Kei Ichikawa, Yasuhiro Hashimoto, Masashi Toyoda, Naoki Yoshinaga, Kazutoshi Sasahara

\author{Qianyun Wu\textsuperscript{$\ddagger \dagger$,*}}
\author{Bruno T. Sugano\textsuperscript{$\dagger$}}
\author{Genta Toya\textsuperscript{$\dagger$}}
\author{Kei Ichikawa\textsuperscript{$\dagger$}}
\author{Yasuhiro Hashimoto\textsuperscript{$\S$}}
\author{Masashi Toyoda\textsuperscript{$\P$}}
\author{Naoki Yoshinaga\textsuperscript{$\P$}}
\author{Kazutoshi Sasahara\textsuperscript{$\dagger$}}

\affiliations{%
  \textsuperscript{$\dagger$}Institute of Science Tokyo\\
  \textsuperscript{$\ddagger$}Hirosaki University\\
  \textsuperscript{$\S$}University of Aizu\\[3pt]
  \textsuperscript{$\P$}The University of Tokyo\\
  \textsuperscript{*}Corresponding author: \href{mailto:wu.qianyun@hirosaki-u.ac.jp}{\upshape wu.qianyun@hirosaki-u.ac.jp}%
}

\makeatletter
\def\shortauthors{Wu et al.}
\makeatother

%\date{\today}

\begin{abstract}
Information cascades on social media are conventionally conceptualized as directed, feedforward branching processes. However, real-world diffusion pathways frequently deviate from pure hierarchical trees due to localized clustering, reciprocal commentary, and multi-wave temporal surges. In this work, we quantify the directional asymmetry and hierarchical structure of empirical information cascades on X (formerly Twitter) using spectral non-normality via Henrici's departure from normality. Analyzing approximately 58,000 cascade networks across diverse topics (including politics, entertainment, natural disasters, etc.), we investigate (1) how non-normality relates to temporal dynamics such as endogenous-like versus exogenous-like patterns and burstiness, (2) whether non-normality is correlated with the peak concentration or overall size of a cascade, (3) whether the overall non-normality of a cascade’s network structure can be predicted from its early stages. We find that non-normality strongly aligns with peak concentration ($peak/N$) rather than overall cascade size, characterizing cascades governed by rapid, asymmetric forwarding. Furthermore, while early-stage structural forecasting ($\le 30\%$ of nodes observed) exhibits expected baseline uncertainty (51\%–72\% accuracy at a $\pm 20\%$ error tolerance), predictability consolidates rapidly during intermediate growth, exceeding 80\% across all dynamic clusters once 50\%–60\% of the network is observed. By identifying the topological and dynamic correlates of cascade structures, this study advances our understanding of information flow and establishes a quantifiable benchmark for forecasting directional diffusion architectures.
\end{abstract} %%%%%%%%%
\maketitle
\renewcommand{\shortauthors}{Wu et al.}

\bigskip

%\noindent \lipsum[1] \cite{1}

$\,$
   
\section{Introduction}

 Information cascades on social media platforms fundamentally shape collective attention, societal sentiment, and public decision-making. In an era characterized by acute information overload, human attention is strictly bounded; consequently, items that achieve broad diffusion disproportionately steer the public agenda. At the individual level, repeated exposure to virally propagated narratives influences affective states and cognitive appraisals, directly impacting personal well-being and downstream behaviors. Collectively, cascade dynamics can reinforce unverified rumors, trigger costly misallocations of public resources, and catalyze offline unrest or social fragmentation. Understanding the generative mechanisms that govern why certain messages dissipate while others undergo explosive virality is therefore a central concern for network scientists, computational social scientists, and policymakers alike.

However, conventional topological metrics, such as degree distributions, centrality measures, and depth or span of networks, predominantly evaluate static graph configurations. As such, they may not fully capture the complexity and dynamics of real-world information diffusion, which can involve localized clustering, reciprocal interactions, and multiple temporal waves. In particular, a method is needed that can capture how directional network structure creates the structural potential for transient amplification and asymmetric information flow.

Non-normality offers such a perspective. In linear dynamical systems, the interaction matrix $A$ governs the evolution of the system, and stability is classically characterized by its eigenvalues. When $A$ is non-normal, i.e., $A A^\top \neq A^\top A$, its eigenvectors are generally non-orthogonal, allowing substantial transient amplification even when all eigenvalues indicate asymptotic stability  \cite{trefethen2005spectra}. While continuous dynamical models establish this principle theoretically, real-world social diffusion presents a compelling structural analog, where asymmetric interaction structures and influence networks can generate collective surges despite remaining spectrally subcritical  \cite{sornette2023non,wang2025dynamic}. Empirical cascades frequently display short-term, explosive bursts despite exhibiting subcritical relaxation profiles. Quantifying the directional asymmetry and feedforward organization of empirical transmission pathways via non-normality provides a structural framework to examine how acute attention concentrations manifest across social networks, echoing how directed architectures control cascade amplification  \cite{sornette2026directedcascades}.

Such transient bursts can have substantial social consequences even when their overall cascade size remains relatively small. A rapid concentration of information diffusion can abruptly increase public attention, accelerate the spread of narratives, and trigger collective responses within a short period of time \cite{sornette2023non,troude2026pseudobifurcations}. Thus, focusing solely on total cascade size may overlook cascades that exert significant short-term influence on the social system. Studying the structural potential for transient amplification therefore provides a complementary perspective for identifying episodes of rapid information mobilization and understanding their potential impact on collective attention and behavior.

Despite recent theoretical progress exploring non-normality in synthetic systems (see Related Work), empirical evidence from large-scale social network cascades remains limited. The few empirical investigations on non-normal networks have focused primarily on specific financial bubbles \cite{sornette2023non} or geophysical shocks \cite{sornette2026nonnormal}. As a result, several fundamental questions about non-normality in real-world information cascades remain unanswered: (1) the extent to which non-normal network structures arise in empirical information cascades, and how their directional and hierarchical organization varies across cascades; (2) how non-normal network structure relates to the temporal characteristics of cascades, including their burstiness, temporal wave patterns, and the distinction between endogenous-like and exogenous-like dynamics; (3) whether non-normality correlates with cascade size and whether eventual non-normality can be predicted from early cascade development.

To resolve these questions, we analyze a massive dataset of cascade events from Japanese-language X (formerly Twitter), comprising approximately 58,000 cascades, which in total contains 1.28 billion interactions (retweets, quotes, and mentions), and 20 million unique users over a two-year observation window from 2024 to 2026. We reconstruct directed, weighted transmission matrices for each cascade and quantify their non-normality using Henrici's departure from normality. We then examine how non-normality relates to the temporal dynamics and size of cascades, and whether eventual non-normality can be predicted from early-stage dynamics.

\section{Related Work}

\subsection{Dynamic Pattern of Information Cascades}

The temporal evolution of information cascades provides important insights into how information spreads and how collective attention rises and decays. Cascades can exhibit rapid bursts, multiple waves, and heterogeneous relaxation patterns that are not captured by their final size alone. %Existing studies used time series analysis to characterize these dynamics. Some studies use stochastic point processes, such as Poisson and Hawkes processes \cite{zhou2021survey}, to model event arrival and distinguish endogenous from exogenous activity \cite{crane2008robust,wu2022classification}, as well as to examine critical and subcritical dynamics characterized by different growth and relaxation patterns. More recently, continuous-time neural models and shape-based time-series clustering \cite{paparrizos2016kshape,cheng2024information,huang2023castemporalgcn} have been developed to capture irregular temporal trajectories. 

To model these dynamics, temporal frameworks such as stochastic point processes \cite{zhou2021survey} and response functions \cite{crane2008robust} are widely employed to characterize event arrival sequences. Within this topic, an established approach is to separate exogenous triggers (spontaneous external shocks) from endogenous propagation (internally driven, self-excited word-of-mouth diffusion) across subcritical and critical relaxation regimes. While supported by an extensive theoretical and empirical literature spanning social media \cite{crane2008robust,wu2022classification}, financial markets \cite{filimonov2012quantifying}, and other complex systems \cite{sornette2005endogenous}, these studies have primarily focused on macroscopic temporal dynamics, leaving the interplay between event timing and the underlying network topology underexplored.

To bridge this gap, recent studies have begun integrating network structure to model and predict the temporal evolution of information cascades. Directions include continuous-time graph neural models \cite{cheng2024information,huang2023castemporalgcn} that capture irregular event trajectories on evolving graphs, as well as multivariate Hawkes processes \cite{sornette2026nonnormal,sornette2026hawkesrank} that explicitly couple edge-level transmission topologies with endogenous self- and cross-excitation mechanisms.

\subsection{Network Structure of Information Cascades}

Prior research characterizing information cascades through structural properties mainly uses cascade size, depth, breadth, branching, and the distribution of nodes across hierarchical levels. 

Studies have examined cascade depth, size, and maximum breadth \cite{vosoughi2018spread,zhao2020fake}, showing that different types of information can exhibit substantially different diffusion structures, for example, false news tends to diffuse deeper than factual news. Other work has introduced width entropy \cite{han2017predicting}, which captures the distribution of cascade width across depths, and demonstrated that structural features such as maximum depth, maximum width, and width entropy can provide useful signals for predicting cascade virality. Another widely used measure is structural virality, commonly quantified by the Wiener Index \cite{zhou2021survey,han2017predicting,vosoughi2018spread} which captures the average pairwise distance between nodes in a cascade. 

Beyond global structural measures, researchers have also examined local branching patterns and the roles of individual nodes. The average branching factor \cite{cheng2018diffusion} of non-leaf nodes has been shown to provide a strong structural signal for distinguishing diffusion protocols, highlighting the importance of local subtree organization in cascade growth. Related work has investigated the structural position of influential users, showing that betweenness centrality \cite{kim2017social}, which captures a node's role as a structural bridge, can be more strongly associated with cascade initiation and influence than degree or closeness centrality. 

Collectively, these studies establish a rich set of topological descriptors for characterizing information cascades. However, these measures predominantly capture the static geometry or local organization of cascade networks and provide limited insight into how directional structural organization may generate transient dynamical effects.

\subsection{Characterizing Flow Asymmetry through Network Non-normality}

Non-normality, as introduced in the Introduction, is particularly relevant to social media because information cascades can take different forms of directed interaction. Some cascades are primarily broadcast-like, in which information flows from a small number of influential users to many others with limited reciprocal interaction, such as the dissemination of news or corrections \cite{zhao2020fake,wu2022classification,wu2025twitter}. Others are more interactive, involving greater reciprocal engagement among users, as often observed in the spread of rumors or memes \cite{zhao2020fake,wu2022classification,wu2025twitter}. These different interaction structures can result in different degrees of non-normality. Theoretically, the degree of non-normality is related to the potential for transient amplification: more strongly non-normal structures can produce greater amplification of activity, even when the underlying network remains asymptotically stable.

Non-normality has been identified across diverse complex systems, including biological \cite{baggio2020efficient}, ecological \cite{asllani2018structure}, economic and socio-economic \cite{sornette2023non}, physical \cite{sornette2026nonnormal,asllani2018structure}, and technological networks. In biological systems \cite{baggio2020efficient}, directed and anisotropic neural, molecular, and cellular networks can exhibit strong non-normality, enabling transient signal amplification and efficient information transmission. In ecological systems \cite{asllani2018structure}, asymmetric predator–prey and species interactions generate non-normal dynamics that can produce substantial transient responses and increase ecosystem vulnerability to perturbations. In economic and socio-economic systems \cite{sornette2023non}, asymmetric influence structures in social trading platforms and online communities can self-organize into hierarchical networks, contributing to transient volatility, bubbles, and viral collective behavior. Non-normality has also been extensively studied in physical systems \cite{sornette2026nonnormal,asllani2018structure}, particularly hydrodynamics, non-Hermitian physics, optics, and seismology, where it provides a mechanism for transient amplification, pattern formation, and responses to external shocks. 

These findings demonstrate the broad applicability of non-normality as a framework for understanding how directional and asymmetric network structures can give rise to transient dynamical responses across complex systems.

However, its application to empirical information cascades on social media remains limited, leaving open questions about how non-normality manifests across real-world cascades and how it relates to their temporal and structural characteristics.

\section{Methods}

\subsection{Data Description}

%Data was retrieved from the X platform via the NTT DATA \emph{Nazuki no Oto} analytical engine, covering the period from April 2024 to February 2026. 
We analyzed X data using the service Nazuki no Oto provided by NTT Data. 
A cascade is defined as a chain of tweets, retweets, quotes, and mentions that share the same root tweet ID. We applied the following criteria to select the candidate cascades:

\begin{itemize}
    \item \textbf{Scale Threshold:} Total cascade volume must satisfy $N \ge 10{,}000$ interactions (including retweets, quotes, and mentions).
    \item \textbf{Non-Promotional Filter:} Commercial advertisements, corporate marketing campaigns, and coordinated promotional cascades were systematically excluded via topic classification to isolate organic collective dynamics from commercial noise.
\end{itemize}

As a result, we included approximately 58,000 cascades encompassing approximately 20 million unique accounts, 58,000 root tweets, 1.1 billion retweets, 100 million direct mentions, and 80 million quote tweets.

Each interaction record contains the following fields (columns): \texttt{tweet id}, \texttt{user id}, \texttt{datetime}, \texttt{referenced tweet id}, \texttt{text}, \texttt{reference user id}, \texttt{referenced type (retweet, quote, or mention)}, \texttt{root id}. 

Cascade threads were reconstructed by indexing all descendants associated with a common \texttt{root id}. 

\begin{figure}[t]
    \centering
    \includegraphics[width=1.0\linewidth]{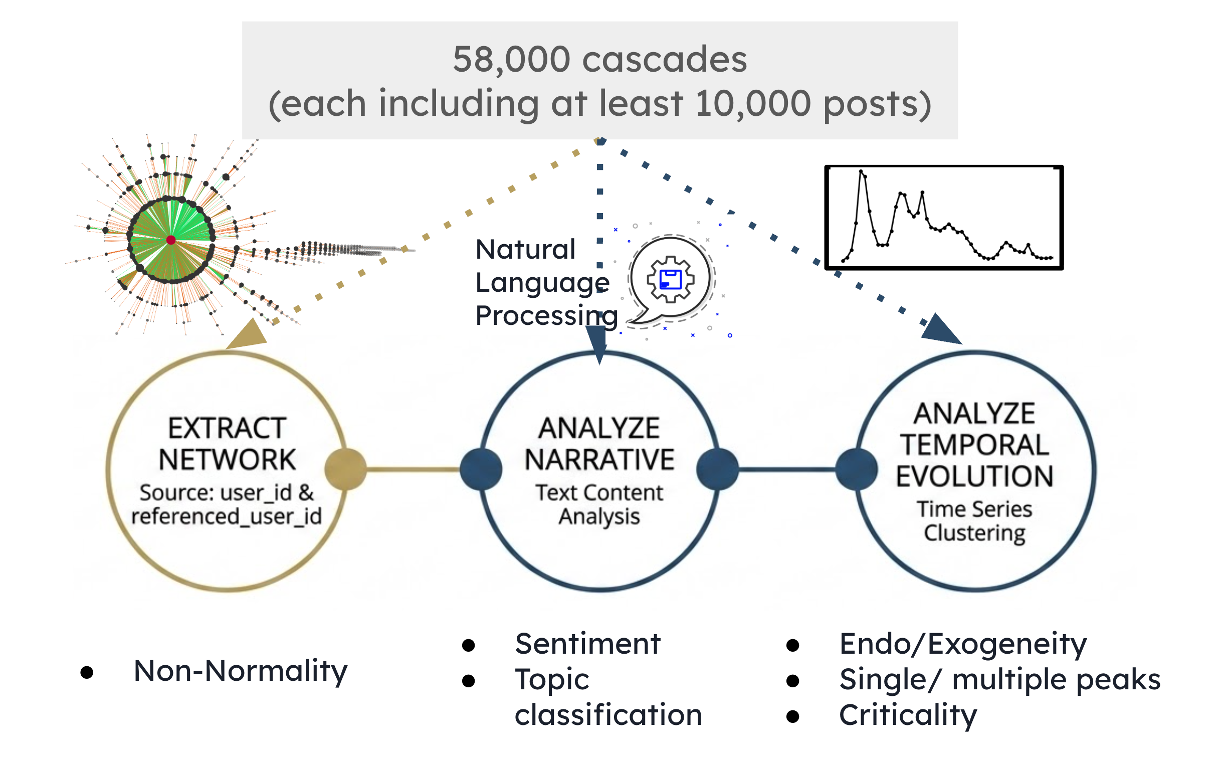}
    \caption{\textbf{Overview of the data analysis framework} }
    \label{fig:overview_analysis}
\end{figure}

The following three subsections describe how we extract narrative and linguistic features, temporal wave features, and network features from these cascade data. Figure~\ref{fig:overview_analysis} illustrates this data processing framework.

%Fig1

\subsection{Narrative and Linguistic Characterization}

\subsubsection{Sentiment Analysis}

Emotional polarity across cascades is evaluated using a cross-lingual transformer pipeline (\texttt{cardiffnlp/twitter-xlm-roberta-base} \cite{barbieri2022xlmt}) \footnote{Available on Hugging Face: \url{https://huggingface.co/cardiffnlp/twitter-xlm-roberta-base}.} fine-tuned for social media text. The classifier assigns each input string a categorical label $y \in \{\text{positive}, \text{neutral}, \text{negative}\}$. Model fidelity was validated against a manually annotated validation set of 100 Japanese tweets randomly sampled from our dataset, yielding an empirical accuracy of $77.0\%$.

To trace sentiment dynamics tractably across the hundreds of millions of text-bearing posts (quotes and mentions), we leverage the heavy-tailed distribution of attention: pure retweets constitute $>90\%$ of total cascade volumes, and diffusion is dominated by high out-degree broadcast nodes. We run sentiment classification directly on the root tweet and on all quoted/mentioned posts within the top $90\%$ of the interaction distribution, providing a tractable proxy of the mainstream emotional trajectory while avoiding computational bottlenecks.

\subsubsection{Topic Classification}

Following the taxonomy of Vosoughi et al. (2018) \cite{vosoughi2018spread}, we categorize cascades into distinct topical domains. We expand the original seven categories (\textit{Politics}, \textit{Urban Legends}, \textit{Business}, \textit{Terrorism/War}, \textit{Science/Technology}, \textit{Entertainment}, and \textit{Natural Disasters}) by introducing two domain-specific classes: \textit{Advertisements \& Campaigns} and \textit{Others}.

Classification is conducted on the root tweet text using the local large language model Qwen 3.0. 
%This model's zero-shot categorization accuracy is tested to be one of the most accurate for classifying Japanese corpora \cite{tian2026benchmarking}. 
Cascades categorized under \textit{Advertisements \& Campaigns} are excluded from downstream dynamical analyses. The query used for text classification is as follows: 

\noindent\fbox{\parbox{\linewidth}{\small\textbf{Prompt:} Classify the following tweet into exactly one of the 9 categories: 1: Politics, 2: Urban Legends, 3: Business, 4: Terrorism/War, 5: Science/Technology, 6: Entertainment, 7: Natural Disasters, 8: Advertisements \& Campaigns, 9: Others. Tweet: ``\textit{[Tweet Text]}''. Return ONLY the single digit number (1--9) with no extra text. Category:}}

\subsection{Temporal Pattern Analysis via K-Shape Clustering}

To capture macroscopic temporal patterns across cascades, we construct an hourly interaction count series $\{N(t)\}$ for each cascade over a uniform time horizon of one week ($t \in \{1, 2, \dots, 168\}\text{ hours}$). $\{N(t)\}$ includes the number of retweets, quotes and mentions. We cluster these count series $\{N(t)\}$ using the \textbf{K-Shape} algorithm, an efficient, scale- and shift-invariant time-series clustering method \cite{paparrizos2016kshape}. 

Unlike distance-based methods that compare point-wise values, K-Shape focuses on the overall shape of time series, making it invariant to differences in magnitude and time shifts. Because K-Shape efficiently computes cross-correlations using the Fast Fourier Transform (FFT), it reduces the pairwise alignment complexity to $\mathcal{O}(m \log m)$ compared to the quadratic $\mathcal{O}(m^2)$ cost of Dynamic Time Warping (DTW).

Given two $z$-normalized time series $\mathbf{x} = (x_1, \dots, x_m)^\top$ and $\mathbf{y} = (y_1, \dots, y_m)^\top$ with zero mean and unit variance, the $k$-Shape algorithm measures morphological similarity using the Shape-based Distance ($\mathrm{SBD}$)~\cite{paparrizos2016kshape}. For an integer shift $s \in [-(m-1), m-1]$, the cross-correlation sequence $\mathrm{CC}_s(\mathbf{x}, \mathbf{y})$ is normalized by the sequence energies to yield the normalized cross-correlation sequence $\mathrm{NCC}_s(\mathbf{x}, \mathbf{y})$:
\begin{equation}
    \mathrm{NCC}_s(\mathbf{x}, \mathbf{y}) = \frac{\mathrm{CC}_s(\mathbf{x}, \mathbf{y})}{\sqrt{\mathbf{R}_{\mathbf{x}\mathbf{x}}(0) \cdot \mathbf{R}_{\mathbf{y}\mathbf{y}}(0)}} = \frac{\mathrm{CC}_s(\mathbf{x}, \mathbf{y})}{\|\mathbf{x}\|_2 \, \|\mathbf{y}\|_2},
\end{equation}
where $\mathbf{R}_{\mathbf{x}\mathbf{x}}(0)$ and $\mathbf{R}_{\mathbf{y}\mathbf{y}}(0)$ denote the zero-lag autocorrelations (equivalent to squared $\ell_2$-norms). The Shape-based Distance $\mathrm{SBD}(\mathbf{x}, \mathbf{y})$ is then obtained by identifying the optimal shift $s$ that maximizes sequence alignment:
\begin{equation}
    \mathrm{SBD}(\mathbf{x}, \mathbf{y}) = 1 - \max_{s} \Big( \mathrm{NCC}_s(\mathbf{x}, \mathbf{y}) \Big).
\end{equation}
By the Convolution Theorem, $\mathrm{CC}_s(\mathbf{x}, \mathbf{y})$ across all $2m-1$ shifts is evaluated efficiently in $\mathcal{O}(m \log m)$ time via the Fast Fourier Transform (FFT). Centroid refinement is formalized as a constrained optimization problem whose exact solution corresponds to the eigenvector associated with the maximum eigenvalue of the phase-aligned cross-correlation matrix.

Similar to standard K-means clustering, K-Shape clustering requires the number of clusters, $k$, to be pre-defined. We determined the optimal value of $k$ using the elbow method evaluated on the within-cluster sum of $\mathrm{SBD}$. The resulting distortion curve exhibits an inflection point at $k=7$, which was selected as the optimal number of clusters.

To characterize the diversity of temporal patterns across cascades, these seven clusters (Fig. 4a) identified via K-Shape clustering are categorized along three dynamical dimensions:

\begin{itemize}
    \item \textbf{Endogenous-like vs. Exogenous-like:} Defined by the growth rate preceding the primary burst. Endogenous-like cascades (C1--C3) exhibit a gradual, multi-step acceleration toward their initial peak, consistent with gradual organic diffusion. Exogenous-like cascades (C4--C7) display an abrupt, steep ascent to the peak, consistent with broadcast events or external informational shocks.
    \item \textbf{Modality:} Evaluated through either single-peak (C1, C4, C5, C7) or two-peak (C2, C3, C6), where secondary peaks reflect recurring bursts of attention, which could be caused by news events or information transmission across communities \cite{almanza2021twin}.
    \item \textbf{Criticality:} Characterized by the relaxation dynamics following the peak of a cascade. For each of the seven cluster-level median time series, we fit the post-peak relaxation curve using either an exponential or a power-law function:

\begin{equation}
    f_{\mathrm{exp}}(t) \sim e^{-(t - t_0)/\tau},
    \qquad
    f_{\mathrm{pl}}(t) \sim (t - t_0)^{-\alpha},
\end{equation}

where the exponential form represents short-memory relaxation, while the power-law form captures slower, fat-tailed relaxation with long-memory characteristics.
\end{itemize}

We fitted an exponential decay function to the post-peak cascade activity using nonlinear least-squares optimization with the Python package \texttt{scipy.optimize.curve\_fit}, using the period from the peak to the end of the cascade.

For power-law fitting, we focus on the initial decay phase following the peak. Specifically, following the procedure in \cite{crane2008robust}, we estimate the relaxation exponent using a least-squares fit on the logarithm of the data over a window beginning 10 time points (10 hours) after the peak. We repeat the fitting procedure over progressively larger windows and select the largest window for which the residuals of the percentage deviation from the fitted curve are consistent with a normal distribution. 

Finally, we compare the exponential and power-law fits using their mean squared error (MSE) and select the model providing the better fit. We classify a cluster as \textit{critical} if its relaxation is better described by a power law with an exponent $\alpha < 1$, indicating slow, fat-tailed decay. In contrast, \textit{subcritical} clusters exhibit either rapid exponential-like relaxation or power-law decay with $\alpha > 1$. Under this criterion, clusters C3 and C7 are classified as critical, whereas C1, C2, C4, C5, and C6 are classified as subcritical. Here, `critical' is used as an operational label for slow, power-law-like relaxation and does not imply that the underlying dynamical system is at a true critical point.

\subsection{Measuring Non-normality: Henrici's Departure from Normality}

%Fig1
\begin{figure*}[t] \centering \includegraphics[width=0.8\textwidth]{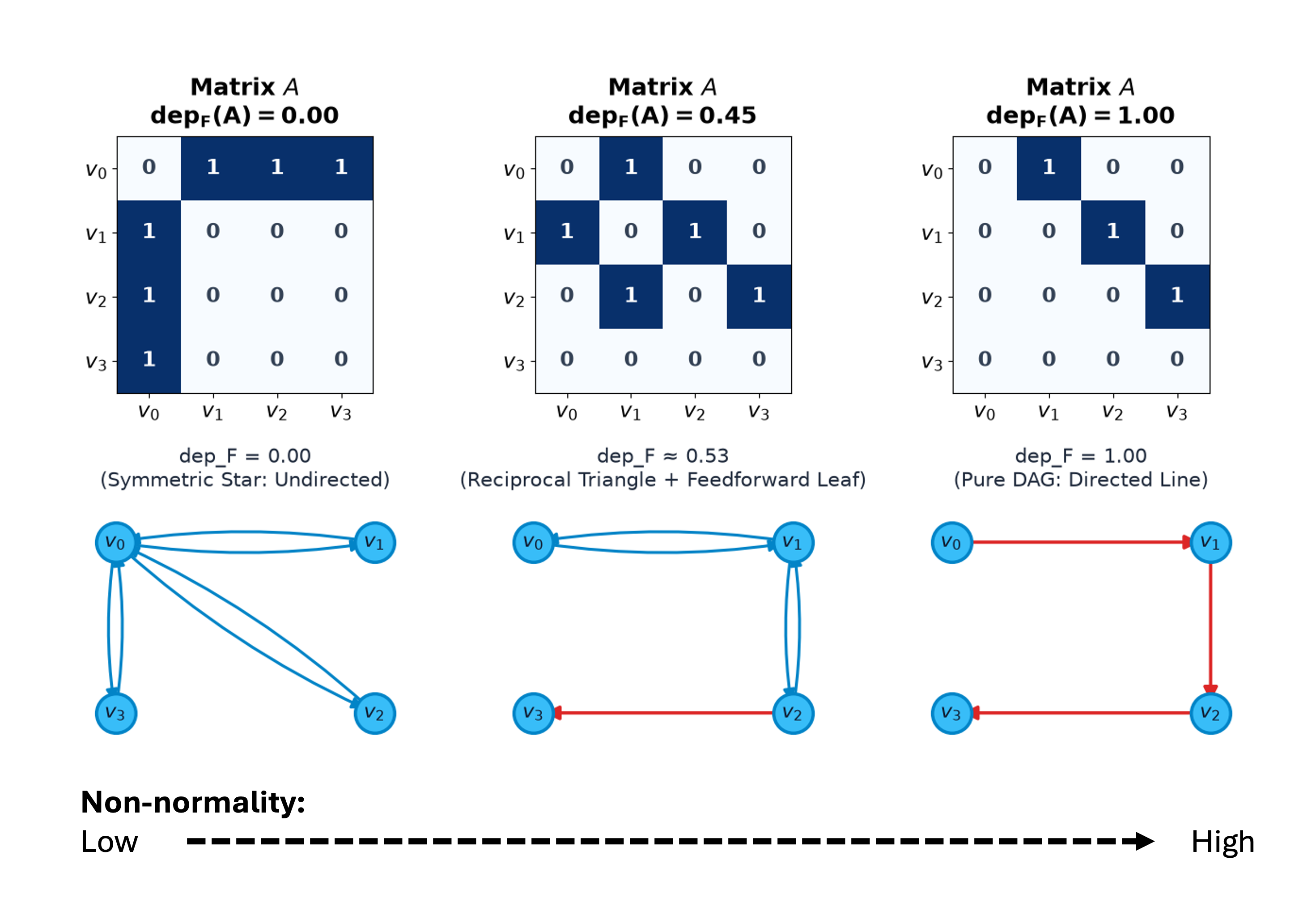} \caption{\textbf{Non-normality as a measure of hierarchy and directedness in a network.}} \label{fig:non_normality} \end{figure*}

For each retained cascade, we construct a directed, weighted interaction graph $G_c = (V_c, E_c, w)$. Nodes $u_i, u_j \in V_c$ denote active user accounts. When user $u_i$ retweets, quotes, or mentions user $u_j$, an asymmetric transmission link is constructed from the source $u_j$ to the receiver $u_i$ ($u_j \to u_i$). The weight $w_{ji}$ represents the cumulative count of directed interactions across the observation window. The corresponding adjacency matrix $\mathbf{A} \in \mathbb{R}^{|V_c| \times |V_c|}$ serves as the basis for our network and spectral non-normality calculations.

To quantify the non-normality of the adjacency matrix $\mathbf{A}$, and thus its departure from an orthogonal eigenmode structure, we compute \textbf{Henrici's departure from normality} \cite{henrici1962bounds,elsner1987measures,trefethen2005spectra,asllani2018structure} $d_F(\mathbf{A})$. Henrici's index is given by:
\begin{equation}
    d_F(\mathbf{A}) = \sqrt{\|\mathbf{A}\|_F^2 - \sum_{i=1}^n |\lambda_i(\mathbf{A})|^2}
\end{equation}
where $\|\cdot\|_F$ is the Frobenius norm. To compare across cascade graphs of different volumes, we use the scale-invariant normalized metric:
\begin{equation}
    \Delta_F = \frac{\sqrt{\|\mathbf{A}\|_F^2 - \sum_{i=1}^n |\lambda_i|^2}}{\|\mathbf{A}\|_F} \in [0, 1]
\end{equation}
Values approaching 1 indicate maximum departure from normality, signaling a strongly asymmetric, non-reciprocal network structure. In particular, any non-empty directed acyclic graph has only zero eigenvalues and therefore attains \(\Delta_F=1\). Figure 2 illustrates three example networks with increasing levels of non-normality, together with their corresponding Henrici's departure from normality. From left to right, non-normality increases as reciprocal connections become less prevalent and the network structure becomes increasingly hierarchical.

%Note that while some previous research \cite{sornette2023non} normalize $d_F(\mathbf{A})$ by $\Vert{}\mathbf{A}\Vert{}_F^2$, which varies inversely with network size across networks of different dimensions. Therefore, we adjusted normalization to $\|\mathbf{A}\|_F$. 

In addition, computing the complete eigenspectrum is computationally costly for large-scale networks. To overcome this bottleneck while preserving spectral fidelity, we compute Henrici's departure using the top $k = 100$ leading eigenvalues obtained via Arnoldi iterations on the sparse adjacency matrix. The spectral magnitude decays rapidly; eigenvalues beyond the 100th become vanishingly small and contribute negligibly to the overall Frobenius norm ($|\lambda_i| \approx 0$), indicating that the truncated approximation closely approximates the full non-normality measure.

\section{Results}

We analyze Japanese-language posts (encompassing original tweets, retweets, mentions, and quotes) collected from the X platform between 2024 and 2026. Individual interactions are mapped to distinct cascades using the root tweet identifier (\texttt{root id}). To capture macroscopic diffusion dynamics and ensure statistical significance across temporal bins, we retain only large-scale cascades satisfying a volume threshold of $N \ge 10{,}000$ interactions. Figure 1 and the Methods section introduce the scope and analytical approach of the study.

\subsection{Distribution of non-normality in social networks}

% Fig. 2
\begin{figure*}[t]
    \centering
    \includegraphics[width=0.9\linewidth]{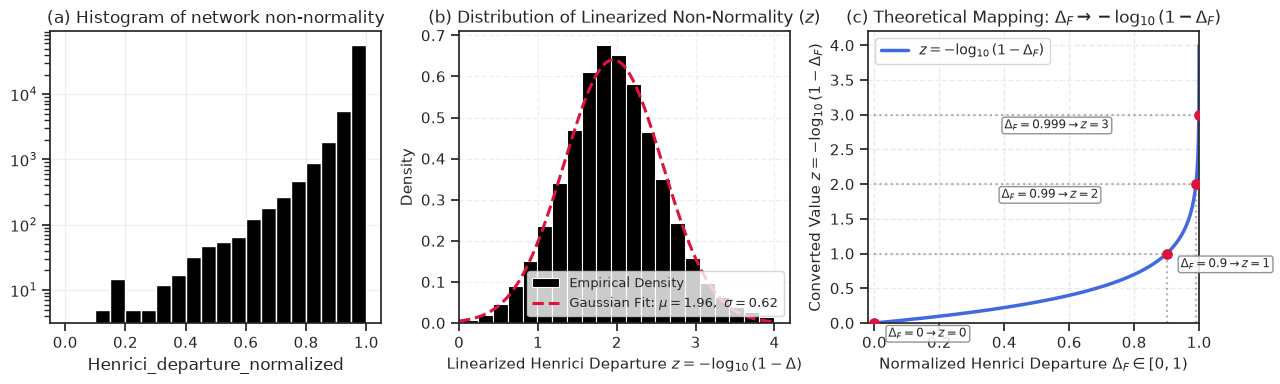}
    \caption{\textbf{Distribution of network non-normality across cascades.}
    (a) Raw frequency distribution of normalized Henrici's departure from normality ($\Delta_F$), exhibiting severe skewness and compression near the theoretical upper bound $\Delta_F \to 1$ on a logarithmic frequency scale.
    (b) Empirical probability density of the transformed non-normality metric $z$, showing that the nonlinear transformation approximately recovers a normal distribution (Gaussian fit: $\mu = 1.96$, $\sigma = 0.62$).
    (c) Theoretical mapping function $z(\Delta_F)$, illustrating how the transformation expands the compressed asymptotic region near $\Delta_F \approx 1$ into a more regular metric space.}
    \label{fig:non_normality_dist}
\end{figure*}

To quantify the degree of non-normality across the cascade interaction networks, we use the normalized Henrici's departure from normality \cite{henrici1962bounds,elsner1987measures,trefethen2005spectra,asllani2018structure}, which has also been applied to non-normal interaction networks in socio-economic systems \cite{sornette2023non} (see the Methods section - Measuring Non-normality: Henrici’s Departure from Normality). 

As shown in Figure 3, the empirical distribution of Henrici's departure is heavily left-skewed, with the vast majority of cascade networks concentrated near the theoretical upper bound of $1.0$ (Fig. 3a). 

This near-maximal non-normality indicates that large cascade networks are strongly dominated by asymmetric, feedforward organization. These topologies are therefore characterized by strongly directional, feedforward pathways with limited cyclic feedback. Consequently, they possess the structural potential for transient amplification associated with non-normal dynamics.

Because the left-skewed distribution can be problematic for linear analyses, such as correlation analysis and linear regression, we transformed the normalized Henrici's departure from normality using the following transformation:

\begin{equation}
z = -\log_{10}(1 - \Delta_F).
\end{equation}

As a result, $z$ approximately follows a Gaussian distribution, with $\mu = 1.96$ and $\sigma = 0.62$ based on a Gaussian fit, as shown in Fig. 3b. This transformation also provides an intuitive mapping of the normalized Henrici's departure from normality (Fig. 3c): $\Delta_F = 0.9$ corresponds to $z = 1$, $\Delta_F = 0.99$ to $z = 2$, and $\Delta_F = 0.999$ to $z = 3$.

%%%

\subsection{Time-Series Shape, Emotion, Topics, and Non-Normality}

%Fig3
\begin{figure*}[!t]
    \centering
    \includegraphics[width=0.9\textwidth]{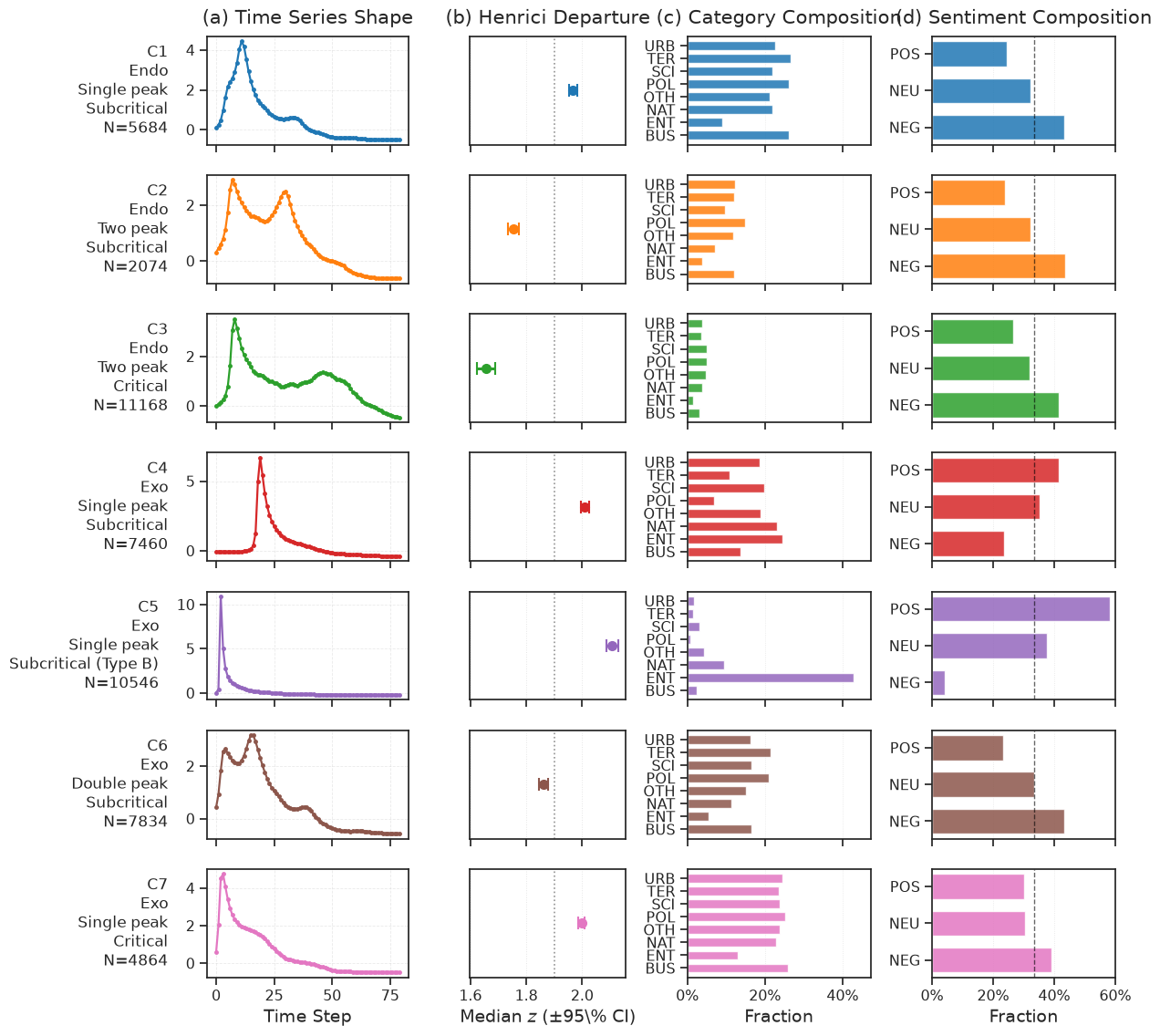}
    \caption{\textbf{Comprehensive overview of the seven cascade clusters.} (a) K-Shape centroid time-series profiles ($k=7$). $N$ in the y-label shows the sample size. (b) Median linearized Henrici's departure from normality ($z$) with bootstrap 95\% confidence intervals (1,000 repeated bootstrap resamples with replacement). (c) Topic category composition (category labels abbreviated to their first three letters) across clusters. (d) Sentiment distribution across clusters. Both (c) and (d) applied two-step normalization to remove the sample size biases.}
    \label{fig:cluster_overview}
\end{figure*}

To understand how non-normality varies across different temporal patterns, we first cluster the time series using K-shape clustering and characterize each temporal-pattern cluster according to three features: (1) endogenous-like or exogenous-like, based on the sharpness of the peak onset; (2) single- or double-peaked; and (3) critical or subcritical, based on the functional form of peak relaxation (see Methods - Temporal Pattern Analysis via K-Shape Clustering). We then calculate the median non-normality for each temporal-pattern cluster and examine the distributions of sentiment and topics across clusters (see Methods - Narrative and Linguistic Characterization).

Figure 4 illustrates a comprehensive characterization of the relationship between temporal dynamics, network non-normality, sentiment, and topical content. Here, we exclude cascades with a normalized Henrici's departure from normality of $\Delta_F=1$, for which the transformation in Eq.~6 is undefined, resulting in a final sample size of approximately 49,000 cascades. In the third and fourth columns, we apply a two-step normalization to account for differences in sample sizes across both categories and temporal clusters. First, for each sentiment or topic category, we normalize the number of cascades within each temporal cluster by the total number of cascades belonging to that category. This removes the effect of differences in the overall frequency of categories, such as the much larger number of neutral posts and Politics-related cascades. Second, we normalize these category-specific proportions across the seven temporal clusters, allowing the relative enrichment or depletion of each category in each cluster to be compared independently of cluster size. This two-step normalization prevents dominant sentiment classes and topic categories from disproportionately influencing the observed patterns.

\begin{figure*}[!b]
    \centering

    % -------- Figure --------
    \includegraphics[width=\textwidth]{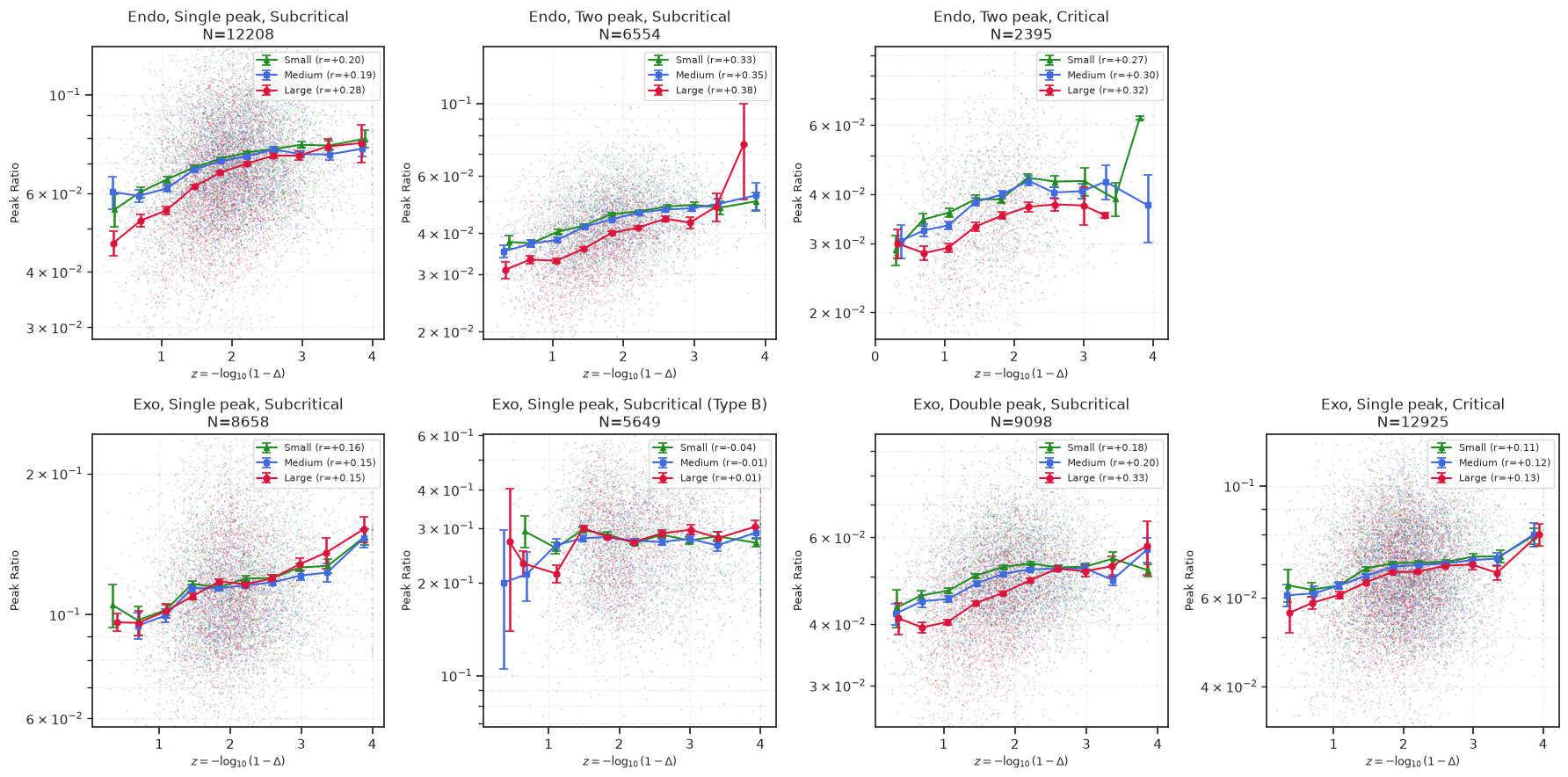}

    \caption{\textbf{Relationship between network non-normality and peak concentration across temporal patterns and cascade sizes.}
    Each panel corresponds to a temporal-pattern cluster characterized by its endogenous/exogenous nature, number of peaks, and criticality. The horizontal axis shows the transformed non-normality measure, $z=-\log_{10}(1-\Delta)$, while the vertical axis shows the peak ratio, defined as the fraction of total cascade activity occurring at the peak. Cascades are further stratified by size into Small (green), Medium (blue), and Large (red) groups with equal sample size. Markers and error bars show the binned peak-ratio values and their variability, respectively (mean $\pm$ standard deviation). The Spearman correlation coefficient \(r\) between non-normality and peak ratio is reported for the individual cascades, while \(r_{\mathrm{bin}}\) denotes the Spearman correlation between non-normality and the binned mean peak ratios.}
    \label{fig:correlation_peak}

    \vspace{0.5em}

    % -------- Table --------
    \begin{minipage}{\textwidth}
        \centering
        \small
        \setlength{\tabcolsep}{5.5pt}
        \renewcommand{\arraystretch}{1.15}

        \captionof{table}{\textbf{Spearman rank correlations between network non-normality and cascade properties.}}
        \label{tab:spearman_correlations}

        \resizebox{\textwidth}{!}{%
        \begin{tabular}{l r ccc ccc ccc ccc}
        \toprule
        & & \multicolumn{3}{c}{$\mathrm{peak}/N$}
        & \multicolumn{3}{c}{$\mathrm{peak}/\mathrm{time}$}
        & \multicolumn{3}{c}{$N$}
        & \multicolumn{3}{c}{$\mathrm{peak}$} \\
        \cmidrule(lr){3-5}
        \cmidrule(lr){6-8}
        \cmidrule(lr){9-11}
        \cmidrule(lr){12-14}
        Dynamic Cluster & $N$ & L & M & S & L & M & S & L & M & S & L & M & S \\
        \midrule
        C1: Endo, Single peak, Subcritical
        & 12,208 & $+0.28$ & $+0.19$ & $+0.20$
        & $+0.21$ & $+0.22$ & $+0.21$
        & $-0.14$ & $-0.01$ & $-0.01$
        & $+0.10$ & $+0.18$ & $+0.19$ \\

        C2: Endo, Two peak, Subcritical
        & 6,554 & \textbf{+0.38} & \textbf{+0.35} & \textbf{+0.33}
        & $+0.28$ & \textbf{+0.30} & $+0.26$
        & $-0.08$ & $-0.07$ & $-0.01$
        & $+0.16$ & $+0.28$ & \textbf{+0.31} \\

        C3: Endo, Two peak, Critical
        & 2,395 & \textbf{+0.32} & \textbf{+0.30} & $+0.27$
        & $+0.18$ & $+0.23$ & $+0.16$
        & $-0.07$ & $-0.02$ & $+0.03$
        & $+0.16$ & $+0.27$ & $+0.26$ \\

        C4: Exo, Single peak, Subcritical
        & 8,658 & $+0.15$ & $+0.15$ & $+0.16$
        & $+0.10$ & $+0.09$ & $+0.12$
        & $-0.03$ & $+0.00$ & $-0.05$
        & $+0.09$ & $+0.14$ & $+0.15$ \\

        C5: Exo, Single peak, Subcritical (Type B)
        & 5,661 & $+0.01$ & $-0.01$ & $-0.03$
        & $-0.04$ & $-0.05$ & $-0.04$
        & $-0.10$ & $-0.03$ & $-0.04$
        & $-0.05$ & $-0.02$ & $-0.04$ \\

        C6: Exo, Double peak, Subcritical
        & 9,098 & \textbf{+0.33} & $+0.20$ & $+0.18$
        & $+0.22$ & $+0.22$ & $+0.23$
        & $-0.11$ & $+0.01$ & $+0.00$
        & $+0.08$ & $+0.18$ & $+0.17$ \\

        C7: Exo, Single peak, Critical
        & 12,925 & $+0.13$ & $+0.12$ & $+0.11$
        & $+0.11$ & $+0.15$ & $+0.14$
        & $-0.10$ & $+0.01$ & $-0.00$
        & $+0.01$ & $+0.12$ & $+0.11$ \\

        \bottomrule
        \end{tabular}%
        }
    \end{minipage}

\end{figure*}

In general, exogenous-like cascades exhibit higher non-normality than endogenous-like cascades, and single-peak cascades demonstrate higher non-normality than two-peak cascades. Based on Fig. 4b, exogenous-like and/or single-peaked time series generally exhibit higher median linearized non-normality ($z>2.0$), whereas endogenous-like and/or double-peaked time series generally exhibit lower median values ($z<1.9$). This is topologically intuitive: endogenous-like bursts and multi-peak dynamics are consistent with organic, word-of-mouth diffusion ~\cite{wu2022classification,almanza2021twin,crane2008robust} and reciprocal debates, fostering denser local clustering and bidirectional links. In contrast, single-peak exogenous-like bursts are consistent with external news shocks or broadcast campaigns, which are associated with directional, feedforward tree structures \cite{zhao2020fake} that maximize operator non-normality. 

Specifically, we observe the following:

%Table1
\begin{itemize}
    \item \textbf{High Non-Normality in Exogenous-like Bursts:} Cluster C5 (exogenous-like, single-peak, subcritical) exhibits the highest non-normality ($z = 2.11 \pm 0.02$). Characterized by an immediate burst dominated by entertainment topics such as release of a new album (Entertainment $> 40\%$) and overwhelmingly positive sentiment (Positive $\approx 60\%$), its propagation is dominated by wide broadcast retweeting with negligible reciprocity, consistent with strongly feedforward organization. Other exogenous-like, single-peaked bursts, such as Cluster C4 ($z = 2.01 \pm 0.01$) and Cluster C7 ($z = 2.00 \pm 0.01$), also exhibit relatively high non-normality. However, their non-normality is lower than that of C5, possibly because their more prolonged discussions, reflected in their fat-tailed temporal profiles, allow for greater feedback and less strongly feedforward propagation.
    \item \textbf{Reciprocity in Endogenous Debates:} Conversely, multi-peak endogenous-like clusters, most notably C3 ($z = 1.60 \pm 0.03$) and C2 ($z = 1.75 \pm 0.02$)—display the lowest non-normality values. These cascades span controversial socio-political categories - Politics, Terrorism/War and Urban Legends) and are heavily enriched with negative sentiment (Negative $> 40\%$). The observed mutual quoting, cross-replies, and modular community segregation are consistent with greater structural reciprocity and lower purely feedforward non-normality, alongside secondary activation peaks. In contrast, single-peak endogenous-like cluster C1 ($z = 1.97 \pm 0.01$) is even higher than the double-peak exogenous-like cluster C6 ($z = 1.86 \pm 0.01$), suggesting that non-normality is not only associated with the origin of the burst, but also relaxation patterns at the tail. 
\end{itemize}

\subsection{Correlation Between Non-normality and Cascade Properties.}

We examined the association between network non-normality and four characteristics of cascade dynamics: relative peak concentration ($\mathrm{peak}/N$), peak speed ($\mathrm{peak}/\mathrm{time}$), total cascade size ($N$), and absolute peak strength ($\mathrm{peak}$) (Table 1), in which the cascades in each cluster is divided into Large, Medium and Small in equal sizes. Across the seven temporal-pattern clusters and three cascade-size tiers, the strongest and most consistent correlations are observed for the relative peak-to-cascade size ratio. In contrast, correlations with total cascade size and absolute peak strength are generally weak, indicating that network non-normality is more closely related to how activity is concentrated within a cascade than to the overall magnitude of the cascade.

This association is particularly pronounced for endogenous-like cascades. For the endogenous, two-peak, subcritical cluster (C2), the Spearman correlation between $z$ and $\mathrm{peak}/N$ is $r_s=0.38$, $0.35$, and $0.33$ for large, medium, and small cascades, respectively. The endogenous, two-peak, critical cluster (C3) shows a similar pattern, with $r_s=0.32$, $0.30$, and $0.27$. The endogenous, single-peak, subcritical cluster (C1) also exhibits positive correlations ($r_s=0.28$, $0.19$, and $0.20$). Thus, the relationship between non-normality and peak concentration is consistently positive across endogenous-like temporal patterns. Exogenous-like cascades, on the other hand, generally exhibit weaker association with cascade dynamics.

Notably, the association with $\mathrm{peak}/N$ tends to become stronger for larger cascades, particularly for endogenous-like patterns. For C2, for example, the correlation increases from $r_s=0.33$ for small cascades to $r_s=0.38$ for large cascades. A similar increase is observed for C3 ($r_s=0.27$ to $0.32$). This size dependence suggests that the relationship between directional network structure and temporal concentration becomes more apparent as the cascade develops and its network structure becomes more extensive.

It is worth noting here that the strongest Spearman correlation is 0.38, corresponding to a weak-to-moderate association. This is expected given that non-normality captures only one structural component of a complex dynamical process. Non-normality characterizes the potential for transient amplification arising from the directional organization of the network, whereas the realistic dynamics of a cascade is additionally influenced by various factors such as external stimuli, user activity, community structure, and stochastic propagation processes. The main purpose of our analysis is therefore not to establish a strong one-to-one relationship between network non-normality and cascade dynamics, but rather to determine whether non-normality provides a systematic structural signal associated with specific temporal characteristics of information diffusion. Therefore, the consistent positive association with peak-to-cascade size suggests that non-normality is more closely related to temporal concentration than overall cascade magnitude, indicating that directional network structure is systematically associated with cascade dynamics alongside other factors.

To reduce the noises in individual cascades and better visualize the relationship between network non-normality and temporal concentration, we plot the peak ratio against \(z\) in Fig. 5, which examines the binned mean peak ratio as a function of non-normality. The resulting trends are substantially clearer, with positive \(r_{\mathrm{bin}}\) values observed across most clusters and cascade-size groups. In particular, the endogenous two-peak clusters (C2 and C3) show pronounced increasing trends, with \(r_{\mathrm{bin}}\) values reaching 0.88–0.98 and 0.70–0.95, respectively. A similar positive trend is observed for the exogenous double-peak cluster C6 (\(r_{\mathrm{bin}}=0.72\)–0.95). The single-peak clusters C1, C4, and C7 also exhibit consistently increasing trends, although the individual-level correlations are weaker. In contrast, Cluster C5 (Type B) shows little systematic relationship, with individual-level correlations close to zero and substantially weaker binned trends.

Although the individual-level Spearman correlations are generally weak to moderate, the binned data reveal a clearer and more consistent positive trend across most clusters and cascade-size groups. This group-level analysis reduces cascade-level variability and highlights the systematic association between higher non-normality and greater peak concentration.

\subsection{Predictability of Final Network Non-normality}

%Fig3
\begin{figure*}[!t]
    \centering
    \includegraphics[width=1\textwidth]{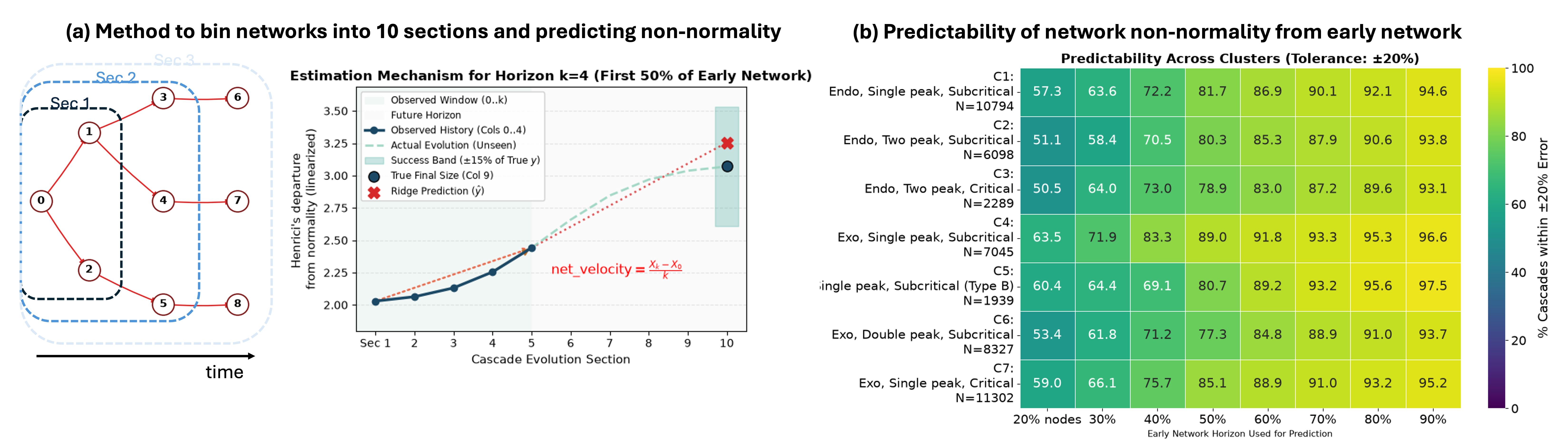}
    \caption{\textbf{Predictability of final network non-normality from early cascade evolution.} (a) Prediction of network non-normality based on initial n\% of nodes. Each cascade is divided into 10 sequential sections, and the evolving non-normality observed over an early portion of the network is used to estimate its eventual value. (b) Prediction performance across the seven temporal-pattern clusters as a function of the fraction of the early network used for prediction. The darkness of color shows the accuracy level, with brighter color representing higher accuracy.}
    \label{fig:predictability}
\end{figure*}

Having established the positive association between peak ratio and network non-normality, we next examine whether the final network non-normality can be predicted from the early development of a cascade (Fig.~6). We divide each cascade into 10 sequential sections in chronological order, with each section containing the same number of nodes. We then progressively increase the fraction of the early network used to predict the final non-normality (Fig.~6a). For example, when the first 50\% of the network is observed, the trajectory of non-normality over the first five sections is used to estimate its subsequent evolution. We employ Ridge regression to characterize this evolution based on the observed trajectory. Specifically, the net velocity of non-normality evolution is defined as $\mathrm{net_velocity}=(z_k-z_0)/k$, where $z_0$ and $z_k$ denote the non-normality at the first and $k$-th observed sections, respectively. The final non-normality is then estimated as $\hat{z}=z_0+\mathrm{net_velocity}\times n$, where $n$ is the total number of sections in the fully developed cascade. We compare the predicted value $\hat{z}$ with the actual non-normality of the fully developed network, $z_{\mathrm{actual}}$. The relative error is defined as $\mathrm{RE}=(\hat{z}-z_{\mathrm{actual}})/z_{\mathrm{actual}}$, and here we assume that a prediction is considered successful when the absolute relative error is within 20\%, i.e., $|\mathrm{RE}|\leq0.20$. For each early-network horizon, we then calculate the fraction of cascades satisfying this criterion to quantify the predictability of eventual non-normality.

Figure 6b shows that the eventual non-normality of a cascade can potentially be predicted from its initial network structure. The heatmap reports the fraction of cascades for which the predicted non-normality achieves a relative error of $\leq 20\%$. Each bin represents the prediction accuracy obtained using an increasing fraction of the initial nodes as input to a Ridge regression model. We can observe that predictability improves substantially as cascades enter intermediate growth phases. By the 40\% horizon, accuracy reaches 69.1\%–83.3\%. Once cascades reach 50\%–60\% of total nodes, prediction accuracy exceeds 80\% across all seven temporal classes. In addition, predictability varies slightly across the seven temporal clusters. C4 (exogenous, single-peak, subcritical) and C7 (exogenous, single-peak, critical) exhibit relatively high predictability even at earlier stages. These results indicate that while early non-normality forecasting ($\le 30\%$) is challenging and constrained by initial diffusion noise, the cascade's directional architecture stabilizes decisively past its midpoint (50\%), enabling reliable inference of eventual network asymmetry long before activity subsides.

Overall, these results demonstrate that the ultimate network non-normality can be predicted from the early structural development of a cascade, indicating that substantial information about the eventual directional organization of the network is already encoded in its initial stages.

\section{Discussion}

This study investigates the relationship between network non-normality and the temporal dynamics of information cascades in social media. By analyzing empirical transmission networks alongside their time-series profiles, we evaluate how directional asymmetry and feedforward organization relate to temporal burstiness, thematic content, and peak concentration.

We characterize cascades according to the shape of their time series, including their origin-like dynamics (endogenous-like versus exogenous-like), peak structure (single versus double peaks), peak strength, and post-peak relaxation behavior associated with criticality. Our results show that network non-normality varies across these time series shapes. In particular, exogenous-like and single-peak cascades, typically associated with positive emotions and Entertainment topics, tend to exhibit higher levels of non-normality. This is possibly because they are more consistent with diffusion initiated by an external stimulus and may therefore develop more concentrated, directional transmission pathways. In contrast, the endogenous-like and multi-peak cascades, which are more observed in negative cascades and politics topics, exhibit lower level of non-normality. This could be associated with more sustained and reciprocal interactions that produce less strongly hierarchical structures.

In addition, our results indicate that non-normality is more strongly correlated with the peak ratio than with either peak activity or total cascade size alone. The peak ratio captures the concentration of cascade activity around its maximum relative to the overall extent of diffusion, and therefore provides a more direct measure of transient burstiness. The positive association between non-normality and peak ratio is consistent with the theoretical expectation that non-normal network structures can support transient amplification even when the underlying system remains asymptotically stable. The association is particularly pronounced for larger cascades and for endogenous-like and multi-peak cascades, which indicates that in more complex cascades, the organization of interactions is more strongly associated with how strongly activity becomes concentrated into transient bursts. 

Finally, our prediction analysis shows that eventual non-normality can be inferred from the early development of a cascade. When approximately 40\% of cascade nodes are observed, a large fraction of cascades can already be predicted accurately. Exogenous, single-peak cascades exhibit particularly high predictability, consistent with their relatively high levels of non-normality. However, prediction accuracy varies across temporal clusters, suggesting that the amount of structural information required to infer eventual non-normality depends on the mode of cascade development. More complex or recurrent cascades may require a larger portion of the network to be observed before their eventual structural organization becomes apparent.

Taken together, these findings highlight the value of studying information cascades by incorporating temporal dynamics and network non-normality. In contrast to conventional cascade measures that characterize either temporal dynamics or network structure in isolation, non-normality provides a complementary measure that links the architecture of information transmission to its transient dynamical behavior. More broadly, the observed association between early network structure and eventual non-normality opens new opportunities for forecasting the eventual structural non-normality of developing cascades and identifying potentially consequential diffusion patterns before their full dynamics unfold.

\section{Limitations}

Despite these findings, several limitations should be acknowledged. 

%{\color{red}Our Henrici-based measure characterizes structural non-normality of the cascade adjacency matrix and does not directly quantify transient amplification of an underlying dynamical operator. Data Size and Selection Bias.}

\textbf{Structural Metric vs.\ Dynamical Amplification.} Our Henrici-based measure characterizes structural non-normality of the cascade adjacency matrix and does not directly quantify transient amplification of an underlying dynamical operator.
Data Size and Selection Bias.

\textbf{Data Size and Selection Bias.} Our dataset is restricted to large cascades containing at least 10,000 posts, excluding smaller diffusion events and potentially introducing selection bias toward big cascades. Consequently, the observed patterns of non-normality may not generalize across the full range of cascade sizes. Future studies should incorporate smaller cascades to examine how non-normality varies across different scales of information diffusion.

\textbf{Unsupervised Clustering Validation.} The temporal patterns identified through K-shape clustering are obtained using an unsupervised approach, for which no ground-truth labels are available to quantitatively evaluate classification accuracy or identify potential misclassifications. Even though we randomly sampled and visually checked the goodness-of-fit and we found that resulting clusters capture distinct temporal characteristics, their boundaries should therefore be interpreted with some caution. Future work could assess the robustness of these temporal classifications using alternative clustering methods, external annotations, or complementary classification approaches.

\textbf{Coarse-Grained Criticality Analysis.} Our classification of critical and subcritical dynamics is based on fitting coarse-grained median time series at the cluster level, rather than fitting power-law or exponential relaxation functions to individual cascade trajectories. While this approach provides a robust characterization of aggregate temporal behavior, it may mask substantial heterogeneity in the decay dynamics of individual cascades. Future research should examine individual cascade decay tails at finer temporal resolution to assess the extent to which the observed criticality patterns hold at the cascade level.

\textbf{Limited Predictive Factors.} Although we identify a positive association between peak ratio and network non-normality, the correlation at the individual-cascade level remains relatively modest, indicating that peak concentration alone does not fully account for variation in non-normality. Other structural, temporal, and content-related factors are not incorporated into the current analysis. Future work could integrate these additional covariates into multivariate predictive models to improve the explanatory and predictive power of non-normality.

\section*{Acknowledgments}
This paper is based on results obtained from a project, JPNP22007, commissioned by the New Energy and Industrial Technology Development Organization (NEDO).

We thank Didier Sornette and Virgile Troude for valuable discussions on non-normal dynamical systems.

$\,$

\bibliographystyle{plainnat}
\bibliography{aaai2026}
%\begin{thebibliography}{99}

%\bibitem{1} Spiegel, M. R. (1981). Theory and problems of Advanced Calculus: Si (metric) edition. McGraw-Hill. 

%\end{thebibliography}

\newpage

\end{document}